\documentclass[a4paper,fleqn]{cas-dc}
\usepackage[numbers]{natbib}

\def\tsc#1{\csdef{#1}{\textsc{\lowercase{#1}}\xspace}}
\tsc{WGM}
\tsc{QE}
\tsc{EP}
\tsc{PMS}
\tsc{BEC}
\tsc{DE}
\usepackage{subcaption} 
\usepackage{caption}
\usepackage{graphicx}
\usepackage{cuted}
\usepackage{flushend}
\usepackage{widetext}
\usepackage{nicefrac}
\usepackage{mathtools}
\usepackage{physics}
\usepackage{mma}
\usepackage{amsmath}
\usepackage{amsfonts}
\usepackage{amssymb}
\usepackage{url}

\begin{document}
\let\WriteBookmarks\relax
\def\floatpagepagefraction{1}
\def\textpagefraction{.001}
\shorttitle{}
\shortauthors{D C et~al.}

\title [mode = title]{Real-Time Electro-Optic Dual Comb Detection of Ultrasound Waves}                      

\author[1]{Debanuj Chatterjee}
\author[1]{Louis Etien}
\author[1]{Simon Boivinet}
\author[2]{Hervé Rigneault}
\author[2]{Thomas Chaigne}
\author[1,3]{Arnaud Mussot}

\address[1]{University of Lille, CNRS, UMR 8523 PhLAM Physique des Lasers Atomes et Molecules, F 59000 Lille, France}
\address[2]{Aix Marseille Univ, CNRS, Centrale Marseille, Institut Fresnel, Marseille, 13013 France}
\address[3]{Institut universitaire de France}

\begin{abstract}
High-resolution ultrasound based imaging techniques like photoacoustic\,(PA) imaging that require fast detection of acoustic waves, are often coupled with an opto-mechanical sensor like a Fabry-P\'erot cavity\,(FPC) for enhanced sensitivity at high frequency. Due to the inherent inhomogeneity of the FPC thickness, the resonance of the cavity can exhibit a spatial distribution, requiring dynamic adjustment of the interrogation wavelength when raster-scanning the sensor with a probe beam. 
To avoid this, we propose in this work the use of an electro-optically modulated dual-comb light source for rapid acoustic wave sensing within a specified bandwidth.
Utilizing a dual-comb vibrometry approach, we demonstrated a proof-of-principle of the technique, with real-time detection of 10 MHz acoustic waves simultaneously with three different teeth\,(separated by 10 GHz) of the dual-frequency comb, achieving a 20 ns temporal resolution. We also investigated the system's sensitivity limit in terms of the strength of the detected acoustic waves, opening new possibilities for ultrafast PA imaging modalities.
\end{abstract}

\begin{keywords}
Dual comb spectroscopy\\
Electro-optic frequency comb\\
Electro-optic sensing\\
\end{keywords}

\maketitle
\section{Introduction}

\begin{figure}[htb!]
\centering
\includegraphics[width=0.99\columnwidth]{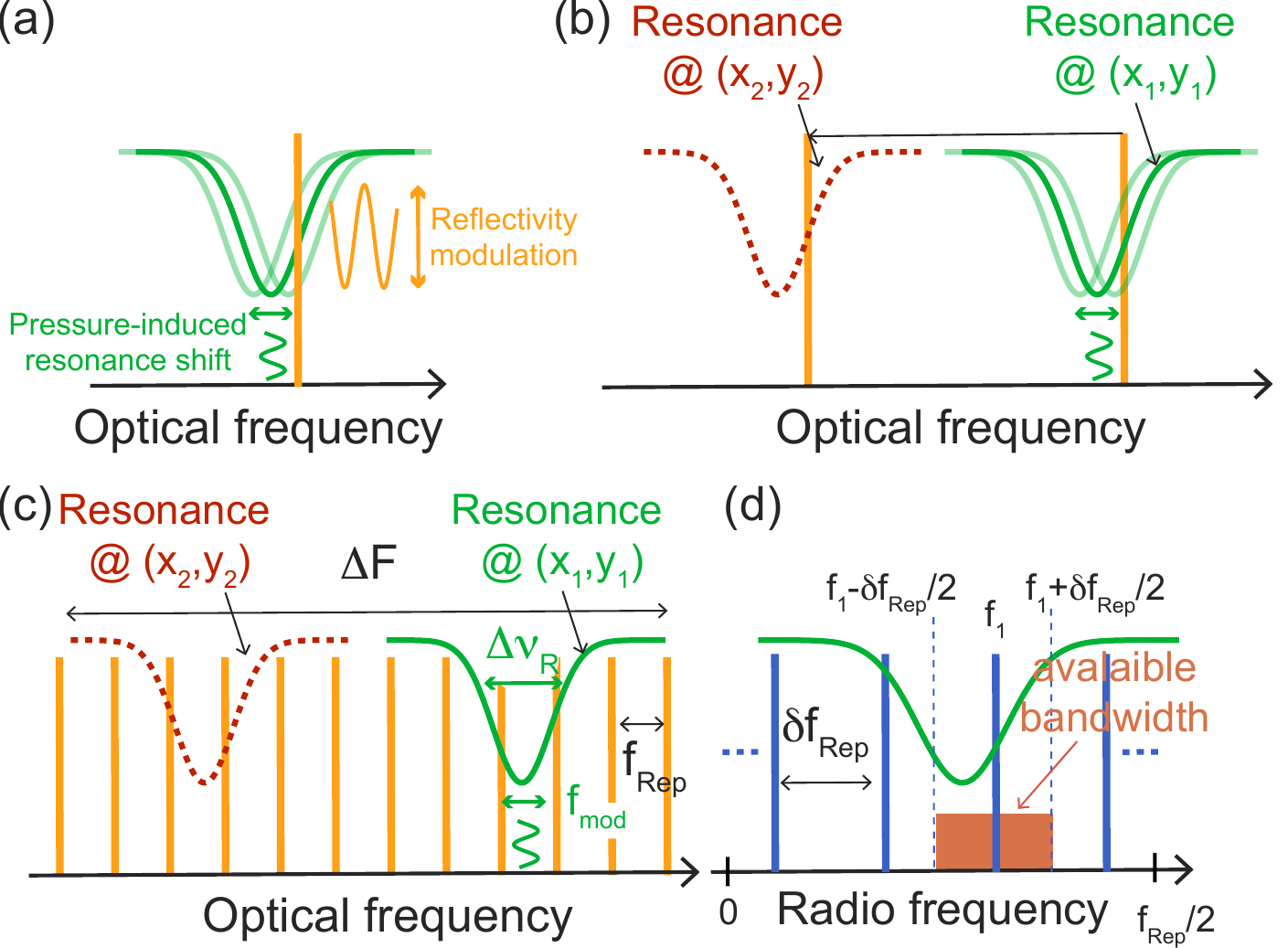}
\caption{Cartoon style scheme illustrating the principle of the DCS analysis for monitoring acoustic waves.  (a) Principle of ultrasound measurement with Fabry-P\'erot cavity\,(FPC): the interrogation wavelength\,(orange line) is centered on the resonance's edge/slope. When pressure waves induce a slight shift of the resonance, the reflectivity of the cavity is  drastically changed. (b) Multiple spectral positions of the resonances corresponding to different spatial positions on the FPC. (c) Tracking multiple resonances of the FPC simultaneously with an optical frequency comb. (d) Available bandwidth\,(dark orange shaded) of modulation for each comb line.} \label{principle}
\end{figure}

Recording ultrasound fields over an extended field of view, in the tens of MHz range, is critical for high-resolution ultrasound-based imaging techniques, such as photoacoustic\,(PA) tomography\,\cite{beard2011biomedical} and\,ultrasound imaging\,\cite{shung_high_2009}.
However, conventional piezoelectric or capacitive micromachined ultrasonic transducer\,(CMUT) arrays face inherent limitations in achieving this goal. Two-dimensional arrays typically exhibit limited bandwidth, and achieving both high frequencies and a large field of view significantly increases system complexity and cost. Optical sensors for ultrasound have emerged as a promising alternative, offering broadband detection and scalable spatial sampling\,\cite{wissmeyer_looking_2018}. Among these, Fabry–Pérot cavity\,(FPC)-based sensors have demonstrated high sensitivity over a broad spectrum and sufficient robustness to enable clinical implementation\,\cite{huynh2024fast}. These sensors detect ultrasound-induced shifts in cavity thickness, which are monitored with a narrowband optical beam. To reconstruct the full ultrasound field, the interrogation beam is raster-scanned across the sensor. However, fabrication-induced inhomogeneities in the polymer layer thickness necessitate dynamic adjustment of the interrogation wavelength\,[Fig.\,\ref{principle}\,(b)], which poses a challenge due to the slow tuning speed of conventional tunable sources. To tackle this problem, various solutions have been proposed, including the use of vertical-cavity surface-emitting lasers\,(VCSEL)\,\cite{zhang_broadband_2006} and broadband sources with fast tunable filters\,\cite{saucourt2023fast}. However, these approaches often involve trade-offs in terms of spectral coverage, tuning speed, and system complexity. Most notably, they prevent parallel interrogation of multiple pixels\,\cite{sievers_fabry-perot_2023,huynh_single-pixel_2019,huynh2024fast} unless exceptional thickness uniformity beyond an optical quality of $\lambda$/100 can be achieved\,\cite{huynh2024fast}. Thus, there is a need for a system capable of covering a large spectrum while enabling ultrafast measurements that can accommodate variations in optical resonance induced by polymer thickness inhomogeneities.

Optical frequency combs\,(OFC) offer a potential solution to this problem. These light sources, consisting of equally spaced and phase-locked laser lines, enable precise measurements\,\cite{cundiff_colloquium_2003, diddams_optical_2020, fortier_20_2019} and ultrafast data acquisition when used in tandem. Two OFCs with slightly different repetition rates\,(also referred to as dual-combs) are often utilized for coherent asynchronous sampling or multiline heterodyne detection, requiring only low-bandwidth photodetectors\,\cite{coddington_dual-comb_2016, picque_frequency_2019}. Rapid dual-comb based techniques have been successfully applied in various fields, including vibrometry\,\cite{teleanu2017electro}, cavity mechanics interrogation\,\cite{long2021electro,long2022high}, laser frequency tracking\,\cite{coddington2011characterizing,giorgetta2010fast}, pulse characterization\,\cite{duran2015ultrafast,ghosh2020fast}, and real-time analysis of chemical samples\,\cite{schliesser2005frequency,coddington2010time, fleisher2014mid, draper_broadband_2019, huh2021time, voumard_ai-enabled_2020, makowiecki2021mid, pinkowski2020dual}. Notably, Ref.\,\cite{long_nanosecond_2024} demonstrated record-high-speed dual comb spectroscopy\,(DCS), achieving a spectral refresh rate of 150 MHz (corresponding to nanosecond time resolution), allowing the capture of fast dynamics in a supersonic jet—however not in real time.

In all-optical PA imaging, OFCs have been utilized for their broad bandwidth and stability\,\cite{rosenthal2012wideband, hazan2019simultaneous, pan2023parallel}, but not in the context of ultrafast detection. Dual-comb sources have also been employed exclusively for excitation in photoacoustic spectroscopy\,\cite{friedlein2020dual, wildi_photo-acoustic_2020, ren2023dual} and imaging\,\cite{stylogiannis_frequency_2022}. However, in all these investigations, even in ultrafast examples, real-time recording of time-resolved dynamics at the sub-microsecond scale, without averaging over multiple interferogram traces has not been achieved. In this work, we demonstrate rapid spectral sampling of PA waves across a large tunable band, accommodating optical resonance variations induced by polymer thickness inhomogeneities using a dual-comb source. Taking advantage of the tunability of dual-combs generated by electro-optic modulators\,(EOM)\,\cite{parriaux_electro-optic_2019}, we optimize both the bandwidth and the repetition frequency for PA applications. In particular, we report real-time measurement of a 10 MHz ultrasound wave detected at a single point on the surface of a FPC. Furthermore, we successfully demonstrated a proof-of-principle dual-comb based fast spectrometer that can effectively interrogate the sensor across its full operating range, offering a scalable and high-resolution solution for ultrasound field mapping.

\section{Principle}

To perform a proof-of-principle for the real-time recording of acoustic waves in the context of all-optical PA imaging, we configured our DCS system to closely resemble a realistic experimental setup. 
Here we probed a FPC sensor made of dielectric mirrors and a 20 $\mu$m thick Parylene C sensing layer on a 1 mm thick glass substrate.  Fabricating ultra-narrow FPCs over surfaces that can be as large as $1\ cm^2$  is challenging, as typical polymer thickness homogeneity of $\lambda/100$ can still result in local resonance shifts up to 300 GHz for thin micrometer-thick layers\,\cite{saucourt2023fast}. For example, in Fig.\,\ref{principle}\,(b) we show an illustration of resonances at two specific positions on the FPC surface, located at ($x_1$, $y_1$)\,(green curve) and ($x_2$, $y_2$)\,(red curve). While standard techniques using a single interrogating continuous-wave\,(CW) laser aligned with the resonance slope can be time consuming due to the need for frequent readjustments, frequency combs can overcome this limitation by covering the entire frequency range at once. We can then define the total bandwidth $\Delta F$ of the comb according to the maximum frequency shifts that resonances can experience across the entire FPC surface\,[Fig.\,\ref{principle}\,(c)]. Another critical parameter for consideration in this pursuit is the repetition rate of the combs $f_{Rep}$. Given that a typical FPC exhibits resonances with 3 dB linewidth of $\Delta \nu_R$, as a rule of thumb, a comb with $f_{Rep}\sim\Delta \nu_R/5$ ensures that at least two comb lines fall on either side of the resonance, maximizing the modulation caused by the resonance displacement. Additionally, the repetition rate difference between the two combs, labeled $\delta f_{Rep}$, must also fulfill the following criteria. Firstly, it is essential to ensure that the sideband from intensity modulation experienced by each comb tooth does not overlap with that of its neighbor, leading to the condition $\delta f_{Rep}>2f_{mod}$\,[Fig.\,\ref{principle}\,(d)]. Secondly, we must respect the standard DCS criterion for avoiding spectral aliasing\,\cite{coddington_dual-comb_2016}, leading to $\delta f_{Rep}<f_{Rep}^2/(2\Delta F)$.  
Thus we have the following constraint on the repetition rate difference $\delta f_{Rep}$: 
\begin{equation}
    2f_{mod} < \delta f_{Rep} < \frac{f_{Rep}^2}{ 2\Delta F}
\end{equation}
From these conditions and the characteristics of the FPC (maximum resonance shift of 300 GHz and a resonance width $\Delta \nu_R\approx$50 GHz), we chose to set the repetition rate $f_{Rep}$ of the combs to 10 GHz, the repetition rate difference $\delta f_{Rep}$ to 50 MHz, and the number of comb teeth to 30, thereby covering a 300 GHz frequency span. Consequently, we had an upper frequency limit of 25 MHz for the acoustic waves to be detected.

\section{Experiment}
\subsection{Setup}
\begin{figure*}[htb!]
\centering
\includegraphics[width=1\textwidth]{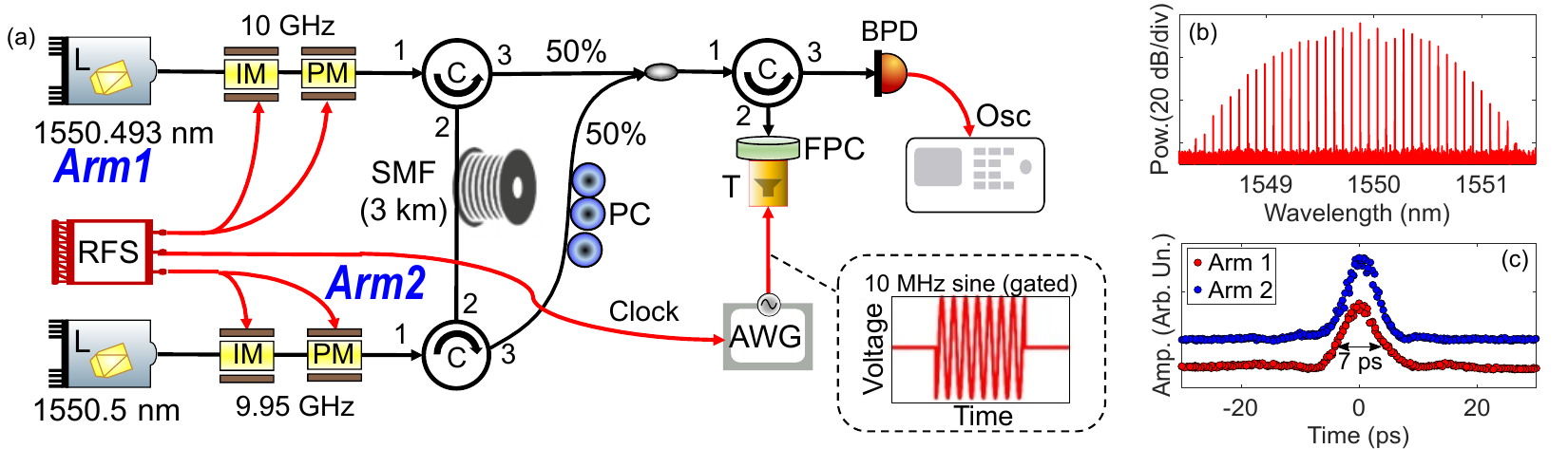}
\caption{(a) Illustration of the experimental setup. A representation of the RF waveform fed to the transducer is shown in dashed box. (b) Optical spectrum of arm 1, recorded with a high resolution Brillouin optical spectrum analyser\,(BOSA) and (c) temporal traces of arms 1 (red) and 2 (blue) recorded with an optical sampling oscilloscope\,(OSO). The different datasets are vertically shifted for visualization. L: laser, RFS: radio frequency synthesizer, AWG: arbitrary waveform generator, IM: intensity modulator, PM: phase modulator, FPC: Fabry-P\'erot cavity, BPD: balanced photodetector, C: circulator, PC: polarization controller, SMF: single mode fiber, T: transducer, Osc: oscilloscope.} \label{scheme_fig}
\end{figure*}

The experimental setup for our experiment is depicted in Fig.\,\ref{scheme_fig}\,(a). Two ultra-narrow linewidth, tunable (over 1 nm bandwidth) lasers (<100 Hz, NKT) with slightly different frequencies (1550.5 and 1550.493 nm respectively), delivering 40 mW each were used as sources in the two arms of the DCS. Each of them passed through an intensity and a phase modulator to generate a chirped pulse train (or EOM comb) in their respective arms. The pulse repetition rates in the two arms (labelled 1 and 2) were 10 GHz and 9.95 GHz\,(meaning $\delta f_{Rep}$=50 MHz) respectively. These pulses were subsequently compressed into 7 ps transform-limited pulses (see Fig.\,\ref{scheme_fig}\,(c), measured with an optical sampling oscilloscope\,(OSO)) after propagation through a 3 km long single mode fiber\,(SMF). To maintain a high mutual coherence between the beams in the two arms of the setup, they were compressed within the same fiber, however propagating in opposite directions to avoid spurious interactions\,\cite{millot2016frequency}. The corresponding spectrum of one of the arms (arm 1) recorded with a Brillouin optical spectrum analyzer\,(BOSA) is shown in Fig.\,\ref{scheme_fig}\,(b), where we see approximately 30 lines with a maximum signal-to-noise ratio\,(SNR) of 50 dB.
After the pulse compression within the SMF, the compressed pulse trains in the two arms were then combined using a 50/50 coupler and directed onto the FPC via a circulator and simple free space optics. The second port of the circulator is imaged onto a $\sim15\ \mu m$ point on the FPC, with a numerical aperture of $\sim 0.03$.  The reflected light from the FPC (about -10 dBm power) was coupled back to the circulator and launched onto a balanced photodetector\,(BPD) (1.6 GHz bandpass) and an interferogram signal was recorded using an oscilloscope with 5 GHz bandpass operating at 20 GSamples/s. 

During the measurement, the thickness of the FPC was periodically modulated with acoustic waves, produced by a focused ultrasound transducer\,(center frequency 10 MHz, bandwidth >70\%, $f$-number = 0.1 Sonaxis), and hitting the back side of the substrate, where the FPC is deposited, through a water-filled cavity. The transducer was driven by bursts of gated sine waves at 10 MHz with a gate duration of 1 $\mu s$ (yielding ten cycles per burst),  and a gate repetition rate of 10 kHz. The RF signal fed to the transducer was generated from a Keysight 81160A arbitrary waveform generator\,(AWG). All electrical signals of the setup (EOM signals and AWG clock) are delivered from low-noise, phase locked multichannel radio frequency synthesizer\,(RFS) from Holzworth. Also, the entire system was all-polarization maintaining except for the compression fiber. Thus, a polarization controller\,(PC) was used in one arm of the DCS to align the polarization states when combining the two arms. 

Since in these experiments we performed ultra-fast recordings ($\sim$20 ns interferogram pulse duration), and applied no averaging over multiple interferogram pulses, we did not observe any mutual coherence degradation despite using two free-running independent lasers for generating the EOM combs in the two arms\,\cite{vicentini2021dual}. 
Furthemore, the use of two independent lasers allowed for achieving a large and tunable central frequency of the RF beating comb\,($\sim$750 MHz), optimized for utilization of the full photodetector bandwidth\,(0 to 1.6 GHz). This strategy allowed us to avoid using an acousto-optic modulator\,(AOM) in our setup, which are often not tunable on a large frequency span, operate at lower frequencies and induce additional losses in the system\,\cite{parriaux_electro-optic_2019}.

\subsection{Results}
\begin{figure}[htb!]
\centering
\includegraphics[width=0.9\columnwidth]{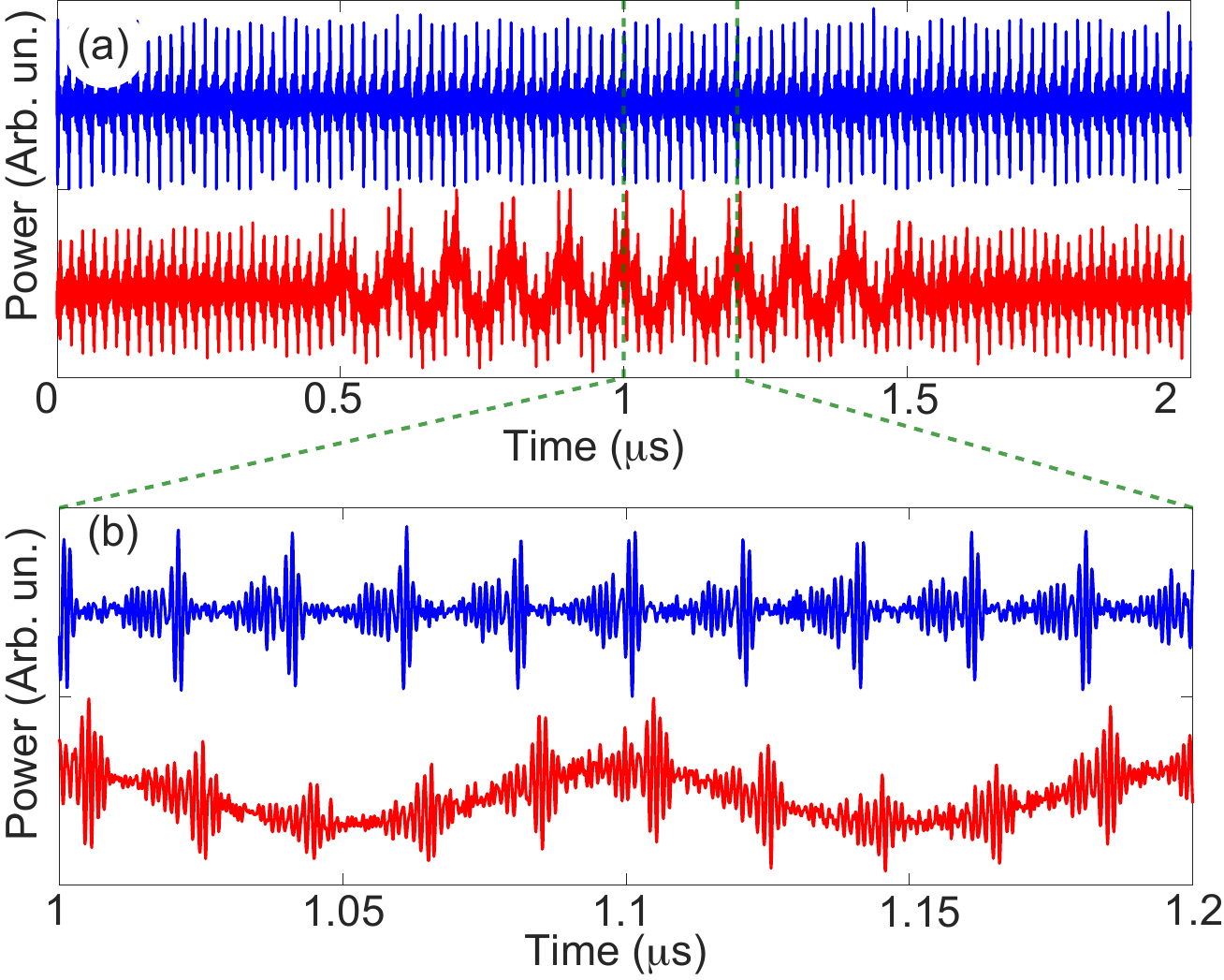}
\caption{Interferograms with (red) and without (blue) modulation. (a) Full trace and (b) zoom.} \label{ifgtime}
\end{figure}

\begin{figure}[htb!]
\centering
\includegraphics[width=1\columnwidth]{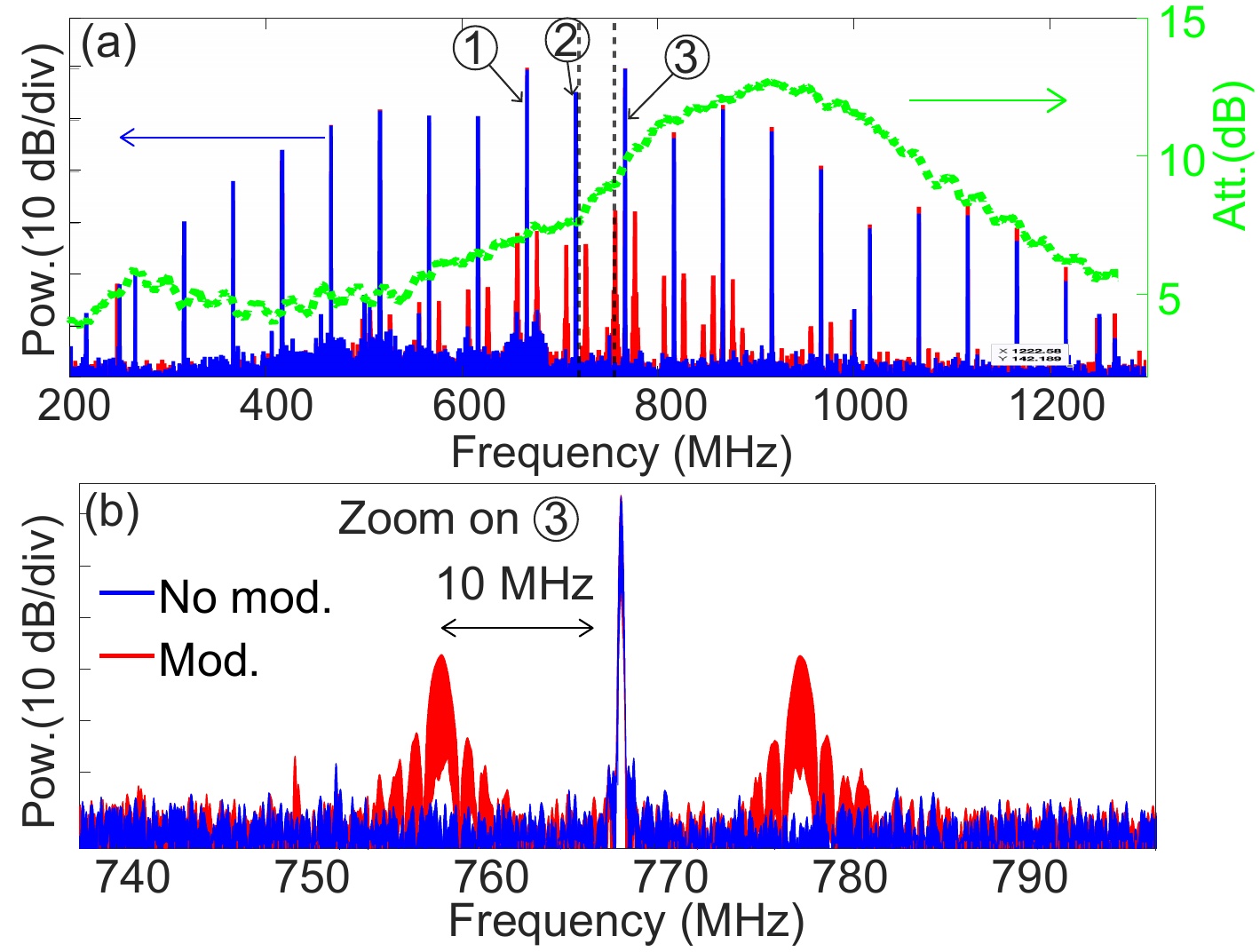}
\caption{Radio frequency spectra with\,(red) and without\,(blue) modulation. (a) Full spectrum and (b) zoom around 768 MHz (tooth 3). The green curve in (a) shows the static reflectivity profile of the FPC rescaled to the RF frequency band. } \label{ifgspectral_fig}
\end{figure}

A sample interferogram recorded with the DCS technique is shown in Fig.\,\ref{ifgtime}. While the blue curves in Fig.\,\ref{ifgtime} show the trace without applying any voltage to the transducer (no FPC thickness modulation), the red curves correspond to a case with peak-to-peak voltage $V_{\text{PP}}$=5 V applied on the transducer (generating a pressure modulation amplitude of approximately 0.5 MPa). 
The resulting intensity modulation is clearly visible on the interferograms (compare red and blue curves in Fig.\,\ref{ifgtime}).
The corresponding RF spectra of the long \,(500 $\mu$s) interferograms (containing 50 acoustic stimulations bursts) are shown in Fig.\,\ref{ifgspectral_fig}. In the static or no modulation case, the reflectivity spectrum of the FPC was recorded using a BOSA shown in green with the right axis in Fig.\,\ref{ifgspectral_fig}\,(a) and the \emph{x}-axis rescaled to the corresponding RF frequency band with a magnification factor $M$=200 and a RF frequency shift of 768 MHz (beat frequency of the two CW lasers). 

From Fig.\,\ref{ifgspectral_fig} we can see that without modulation, a standard RF spectrum is obtained. It is centered at 768 MHz, the beat frequency of the two CW sources and made of spectral lines separated by the repetition rate difference of 50 MHz. When the modulation is applied, the teeth which are located along the slope of the FP resonance experience the most significant intensity modulation at 10 MHz, as can be seen in some of the comb teeth in Fig.\,\ref{ifgspectral_fig}\,(red curves). As an example, we selected the 3 most significant teeth (labeled 1, 2 and 3) in Fig.\,\ref{ifgspectral_fig}\,(a) located at 668 MHz, 718 MHz and 768 MHz. For the teeth located at lower/higher frequencies, the modulation is still visible but weak due to the low resonance slope and/or the low SNR of the teeth.
A zoom on tooth 3 is shown in Fig.\,\ref{ifgspectral_fig}\,(b). Two side lobes appear symmetrically around this teeth, shifted by 10 MHz, which corresponds to the frequency of the acoustic wave. 
\begin{figure}[htb!]
\centering
\includegraphics[width=0.9\columnwidth]{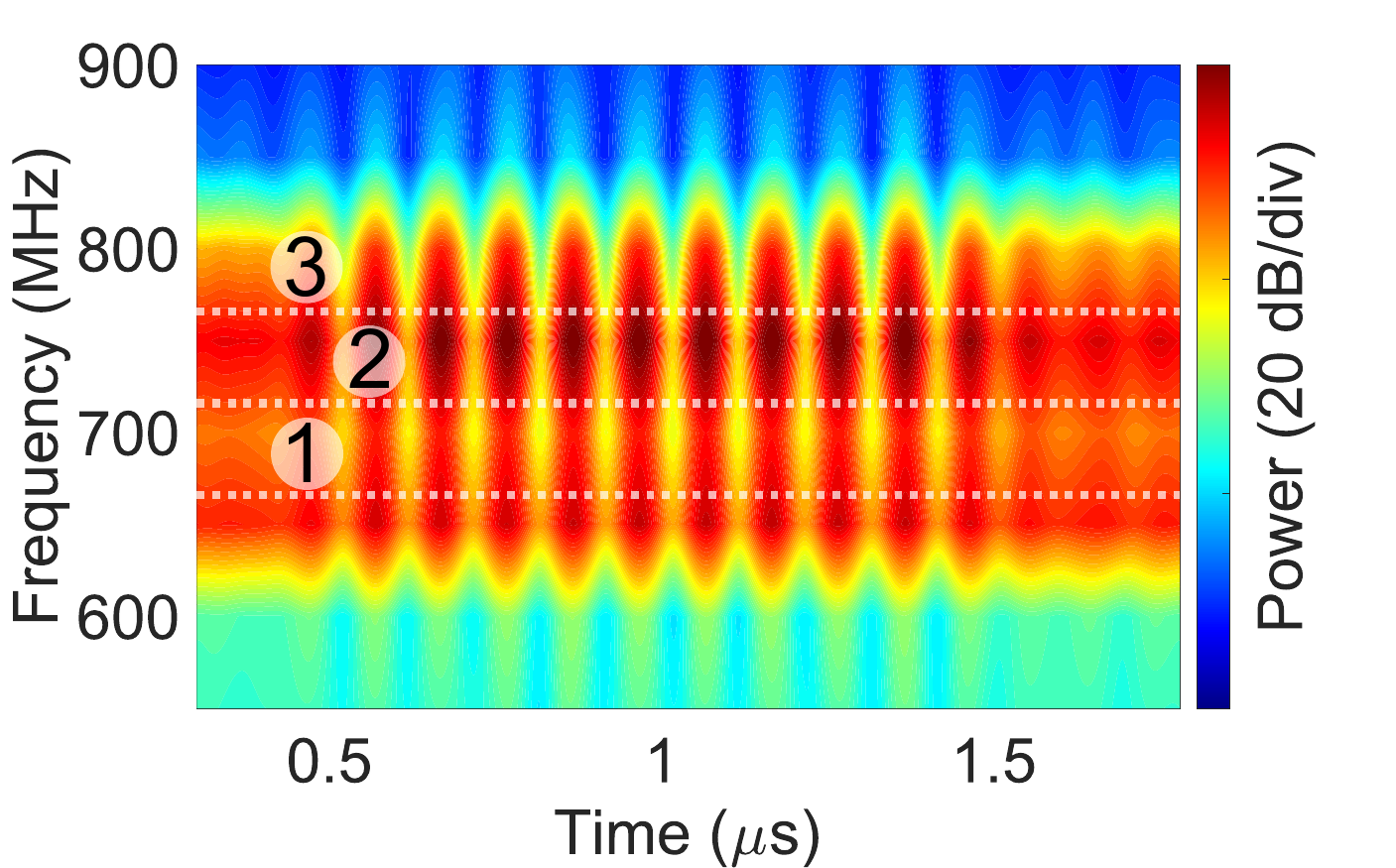}
\caption{Spectrogram calculated from Fig.\,\ref{ifgtime}\,(a). Teeth 1, 2 and 3 are indicated with white dashed lines.} \label{spectrogram_fig}
\end{figure}
The side lobes exhibit a sinus cardinal shape due to the gated modulation applied to the transducer. From this shape, we retrieve the gate duration of 1 $\mu$s, indicated by the zeros at ±1 MHz from the lobe's maximum.
\begin{figure}[htb!]
\centering
\includegraphics[width=0.9\columnwidth]{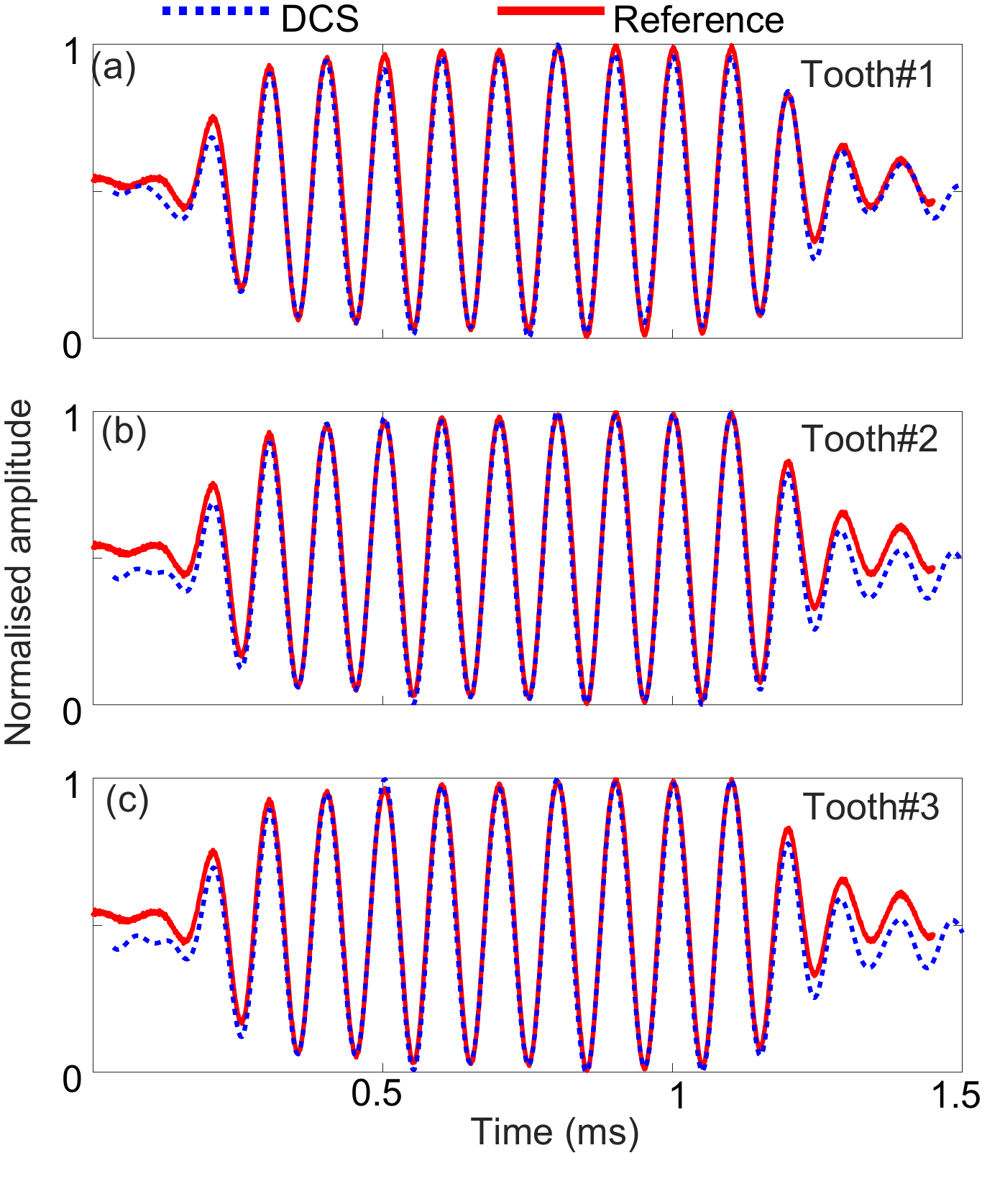}
\caption{Normalized retrieved signals from DCS\,(dashed blue) and reference\,(red solid) measurements. (a)-(c) for teeth 1-3 from Fig.\,\ref{ifgspectral_fig}(a) respectively.} \label{mainresult_fig}
\end{figure}

To extract the amplitude modulation of each tooth, the interferograms were analyzed using a short time Fourier transform\,(STFT) technique. A Hamming-shaped time window with a duration of 1/$\delta f_{rep}=20 $ ns was applied, which give the temporal resolution of our system. The obtained spectrogram is represented in Fig.\,\ref{spectrogram_fig}. The teeth labeled 1,2 and 3 in Fig.\,\ref{ifgspectral_fig}\,(a) are marked with dashed white lines in the spectrogram. We clearly see the modulation experienced by the different teeth in Fig.\,\ref{spectrogram_fig}. The intensity modulation imprinted on the different tooth were extracted from the spectrogram by applying a spectral filter of 30 MHz bandwidth centered on the corresponding teeth. The time evolution for each tooth (from 1 to 3) is shown in Fig.\,\ref{mainresult_fig}\,(a)-(c), respectively. These traces are compared to a reference signal corresponding to the modulation experienced by a single CW laser located on the slope of the cavity resonance, whose temporal modulation is directly detected by a photodetector. This corresponds to the conventional interrogation scheme of such a FPC. From Fig.\,\ref{mainresult_fig}, we see that for each tooth, the agreement between the reference\,(red solid) and the DCS-retrieved\,(blue dashed) is excellent, demonstrating that our system can monitor and reconstruct fast acoustic waves in real-time at frequencies up to 10 MHz.

\begin{figure}[htb!]
\centering
\includegraphics[width=0.9\columnwidth]{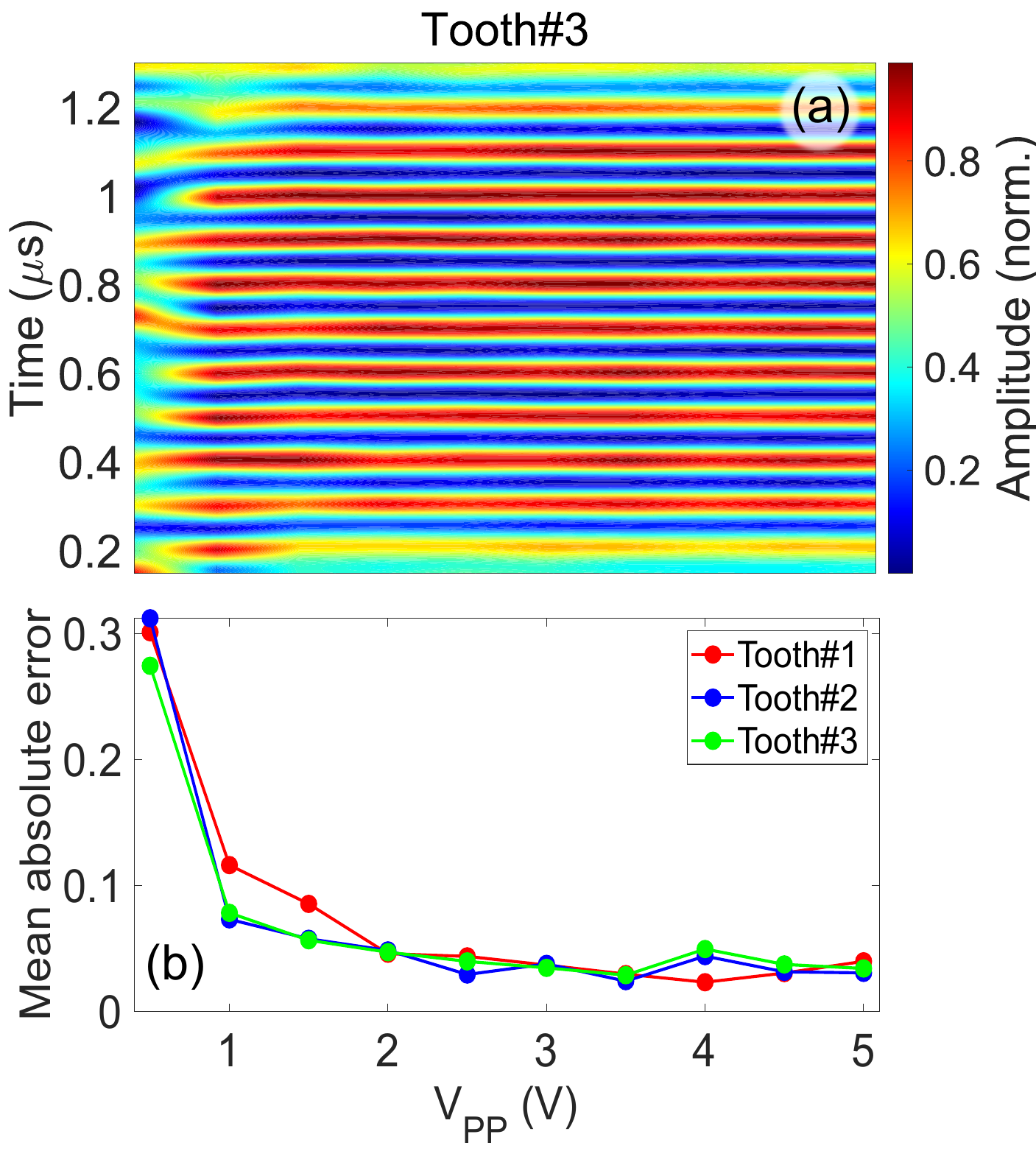}
\caption{(a) False color plot of normalized amplitude of the retrieved signal from DCS (for tooth 3) as a function of time and $V_{\text{PP}}$ applied on the transducer. (b) Evolution of the mean absolute error (with respect to reference measurement) as a function of transducer's $V_{\text{PP}}$, for teeth 1\,(red), 2\,(blue) and 3\,(green). } \label{accuracy}
\end{figure}

To gain a deeper insight into the limitations of the method, Fig.\,\ref{accuracy} illustrate the evolution of the accuracy of the monitored signal as a function of the applied peak-to-peak voltage $V_{\text{PP}}$ on the transducer, representing the amplitude of the acoustic waves\,[see Fig.\,\ref{accuracy}(a)]. 
The accuracy is defined as the mean absolute error (deviation between DCS and reference measurement) between the normalized retrieved signal and the normalized reference signal (red and blue curves in Fig.\,\ref{mainresult_fig}) over the time window between 0.2 and 1.2 $\mu$s (middle of the gate sine function).
By lowering $V_{\text{PP}}$, the amplitude of the acoustic waves decreases, leading to a drop in the accuracy of retrieval of the acoustic signal. This is evident from Fig.\,\ref{accuracy}\,(a), where we clearly see the consistent 10 MHz retrieved modulation signal for $V_{\text{PP}}$>1 V. However, below 1 V, the retrieved signal is distorted, and the corresponding mean absolute error as shown in Fig.\,\ref{accuracy}\,(b) shoots up. 

\subsection{Discussion}
In this work, we recorded an acoustic wave modulated at 10 MHz at a fixed position on a FPC. Due to experimental constraints, repositioning the laser beam to record measurements at different locations was not possible, preventing compensation for shifts in the FP resonance caused by fabrication imperfections. In order to realize a fair proof-of-concept, we developed a dual-comb system with characteristics that enable measurements even when the FPC resonance shifts over a 300 GHz frequency span, which falls within the range of typical fabrication variations. The modulation was detected using spectral teeth near the center of the frequency comb; however, given the relatively flat spectral profile of the comb, similar results are expected for other teeth, corresponding to different positions of the reflective absorption band.

\section{Conclusion}
In this work we have experimentally demonstrated the potential of EOM based dual-comb light sources for real-time opto-mechanical sensing especially in the context of fast photoacoustic imaging. In particular, as a proof-of-principle, we measured a high frequency acoustic wave at 10 MHz through time-frequency analysis of the DCS interferogram with a 20 ns temporal resolution and single-shot recording. These results open new possibilities in frequency comb based metrology techniques that demand simultaneous access to high spectral and temporal resolution. Further improvements of the system in terms of the spectro-temporal resolution might be inspired from compressive sensing\,\cite{kawai2021compressive}, time-programming\,\cite{chatterjee2025sensitivity} or other multiplexing techniques\,\cite{coddington2010coherent} used in DCS technologies.

\section*{Acknowledgments}
DC, LE and AM acknowledge Agence Nationale de la Recherche (Programme Investissements d’Avenir, FARCO projects), Ministry of Higher Education and Research; Hauts de France Council (GPEG project); European Regional Development Fund (Photonics for Society P4S). Support from project FRESCOS (Marie Curie Postdoctoral Fellowship) is acknowledged by DC and AM. This work was also supported by the European Research Council (starting grant ALPINE-101117471 to TC).

\printcredits

\bibliographystyle{ieeetr}

\bibliography{versionArXiv}

\end{document}